\documentclass[aps,twocolumn,showpacs,preprintnumbers,amsmath,amssymb,floatfix]{revtex4}

\usepackage{graphicx}
\usepackage{dcolumn}
\usepackage{bm}
\usepackage[german,american]{babel}

\begin{document}


\title{The Origin of the Fragile-to-Strong Crossover in Liquid Silica as Expressed by its Potential Energy Landscape}

\author{A. Saksaengwijit}
\author{J. Reinisch}%
\author{A. Heuer}%

\affiliation{
Westf\"{a}lische Wilhelms-Universit\"{a}t M\"{u}nster, Institut f\"{u}r Physikalische Chemie\\
and International Graduate School of Chemistry\\
Corrensstr. 30, 48149 M\"{u}nster, Germany }

\date{\today}

\begin{abstract}

The origin of the fragile-to-strong crossover in liquid silica is
characterized in terms of properties of the potential energy
landscape (PEL). Using the standard BKS model of silica we
 observe a low-energy cutoff of the PEL. It is
shown that this feature of the PEL is responsible for the
occurrence of the fragile-to-strong crossover and may also explain
the avoidance of the Kauzmann paradox. The number of defects, i.e.
deviations from the ideal tetrahedral structure, vanishes for
configurations with energies close to this cutoff. This suggests a
structural reason for this cutoff.
\end{abstract}

\maketitle

Understanding the properties of liquid silica is of utmost
importance due to its enormous technological and scientific
relevance. In the Angell \cite{Angell:1991} classification scheme
it is a very strong glass, i.e. displaying an Arrhenius
temperature dependence of its transport observables like viscosity
or diffusivity. Furthermore one may expect that as a typical
network-forming system it shares properties with other
network-formers like water. Indeed, recently it has been pointed
out that anomalies like negative thermal expansion, isobaric heat
capacity minima and isothermal compressibility minima occur in
water as well as in simulated liquid
silica\cite{Vollmayr:1996,Voivod:2001,Shell:2002}.

The experimental oxygen diffusion constant
 $D(T)$ behaves like
$D_0 \exp(-E_a/k_B T)$ with $D_0 = 2.6$ cm$^2$/s
\cite{Mikkelsen:1984}. For a microscopic hopping time of $\tau_0 =
10^{-13}$ s and a typical length scale of a=2 \AA \, one would
expect $D_0 \approx a^2/(6\tau_0) \approx 10^{-3}$ cm$^2$/s. This
discrepancy of some orders of magnitude could be rationalized if
for temperatures above the region, which is accessible in most
experiments, the diffusivity $D(T)$ would show deviations from the
Arrhenius behavior. Indeed, from experiments \cite{Hess:1996} as
well as from simulations \cite{Horbach:1999} this so-called
"fragile-to-strong" crossover (FSC) has been found for silica in a
temperature range around $T_{FSC} \approx 3500$ K. A similar
scenario has been discussed for water \cite{Angell:1983} and
BeF$_2$ \cite{Hemmati:2001}. It has been speculated that the
occurrence of a FSC is directly related to the presence of
polyamorphism \cite{Starr_Sciortino:1999,Voivod2:cond}.

In recent years the analysis of the potential energy landscape
(PEL) became an indispensable tool to elucidate the thermodynamics
and dynamics of many different supercooled liquids
\cite{Stillinger1:1982,Sastry:1998,Sciortino:1999,Buechner:1999,Scala:2000,
Debenedetti:2001,Wales:2001,Sastry:2001,Nave_Sciortino:2002,Shell:2003,Doliwa_hop:2003,
Voivod1:2001,Voivod2:cond}. The dynamics of the total system is
represented as the dynamics of a point in the high-dimensional
configurational space. In particular, the thermodynamics can be
expressed in terms of the distribution of the inherent structures
(minima of the PEL) \cite{Sciortino:1999,Voivod:2000}. It was also
possible to express the long-range dynamics, i.e. the diffusivity,
in terms of PEL-characteristics
\cite{Nave_Sciortino:2002,Doliwa_tau:2003}. Thus the phenomenon of
the FSC should be reflected by properties of the PEL. Actually,
for liquid silica it has been shown that around $T_{FSC}$ the
temperature dependence of the average inherent structure energy
and, correspondingly, the configurational entropy $S_c(T)$ shows
an inflection point \cite{Voivod2:cond}. This has direct
consequences for the Kauzmann paradoxon. Extrapolating the
$S_c(T)$-dependence for $T > T_{FSC}$ to lower temperatures one
would obtain $S_c(T_K) =0$ for a finite temperature $T_K$, the
Kauzmann temperature. The inflection point around $T \approx
T_{FSC}$ suggests that $S_c(T)
> 0$ for all temperatures. Thus it is speculated that "the FSC and
the Kauzmann paradox are fundamentally interrelated phenomena"
\cite{Voivod2:cond}. In this work we will show that this is indeed
the case and suggest the underlying reason for both properties.

For the modelling of silica we use the standard BKS potential
\cite{BKS:1990}. As compared to the standard choice we employ a
somewhat larger cutoff-radius of 8.5 \AA \, for the short-range
interaction to avoid energy drifts. Simulations are performed in
the NVE ensemble with periodic boundary conditions, using a
density of $\rho = 2.30$ g/cm$^3$. The lowest simulated
temperature was 2800 K. In recent years the concept of metabasins
(MB) has been introduced as an appropriate way of coarse-graining
the configuration space
\cite{Stillinger:1995,Buechner:2000,Middleton:2001}. On a
qualitative level, adjacent inherent structures between which the
system performs back- and forth jumps are merged together to one
MB (see \cite{Doliwa_hop:2003} for a precise definition). The
energy $e$ of a MB is defined as the energy of the inherent
structure with the lowest energy. Whereas for the thermodynamics
the distinction between inherent structures and MBs is not
relevant, the relation to the dynamics is much simpler in terms of
MBs. For a binary Lennard-Jones system it could be shown that the
average residence time in a MB but not that in an inherent
structure is directly proportional to the inverse diffusion
constant \cite{Doliwa_tau:2003}.

To characterize the PEL it is crucial to analyse rather small
systems with periodic boundary conditions \cite{Doliwa_hop:2003,
Wales:2003}. Otherwise, interesting aspects may be simply averaged
out, as shown below. For this purpose one may look for the minimum
system size without any significant finite size effects. For a
binary Lennard-Jones system it could be shown that starting from
circa 65 particles no finite-size effects are present for the
range of temperatures accessible to computer simulations
\cite{Doliwa_finite:2003}. For BKS silica it is known that the
diffusion constant becomes constant only for system sizes of more
than 1000 particles \cite{Horbach_finite:1996}. We have calculated
the incoherent scattering function $S(q_{max},t)$ for system sizes
ranging from 60 to 1000 at $T \approx T_c$. It turns out that they
all have identical non-exponential relaxation behavior in the
$\alpha$-regime except for simple scaling (e.g.
$\tau_\alpha(N=99)/\tau_\alpha(N=1000) \approx 3$); see inset of
Fig.1. This is a first hint that the transport behavior is
identical for all system sizes in this range. In what follows we
report simulations for $N=99$ (denoted as BKS99).

The temperature dependence of the oxygen diffusion constant $D(T)$
of BKS99 is shown in Fig.1a. It displays the same FSC as the
macroscopic system with Arrhenius behavior below 3500 K and a very
similar low-temperature activation energy (4.9 eV). In Fig.1b we
show the average energy of MBs $\langle e(T) \rangle$ and inherent
structures $\langle e_{IS}(T) \rangle$. Furthermore we include the
inherent structures energies from Ref. \cite{Voivod2:cond} where a
much larger BKS silica system ($N=1332$) with slightly different
cutoff conditions and nearly the same density ($\rho = 2.31$
gcm$^{-3}$) has been simulated. The temperature dependence is
nearly identical for all three curves and has a change in slope
around $T_{FSC}$. We note in passing that the temperature
dependence of the dynamic heterogeneities in real space does not
show any anomalies around $T_{FSC}$ \cite{Vogel:2004}.

\begin{figure}
\includegraphics[width=8.6cm]{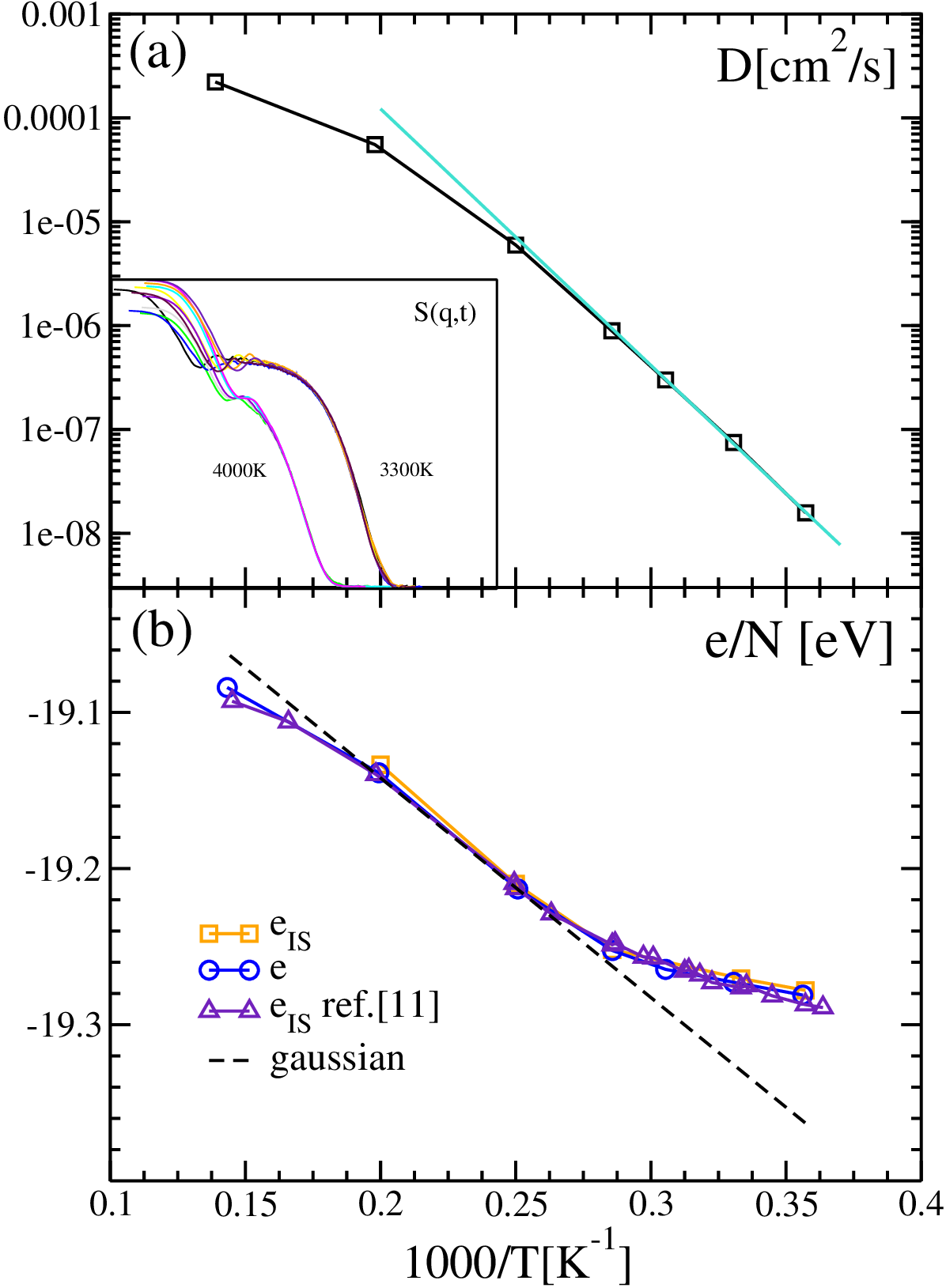}
\caption{\label{d2eff} (a) Temperature dependence of the oxygen
diffusion constant for BKS99. In the inset the incoherent
scattering function $S(q_{max}=1.7$\AA$^{-1},t)$ is shown at $T =
3300$ K and 4000 K for system sizes between 60 and 1000 and after
scaling along both axes. (b)Temperature dependence of the average
energy of MBs $\langle e(T) \rangle$, the average energy of
inherent structures $\langle e_{IS}(T) \rangle$. Included are data
from Ref.\cite{Voivod2:cond}, shifted by 0.06 eV. The broken line
describes the gaussian approximation with parameters obtained from
Fig.2.}
\end{figure}

So far we have reproduced the {\it temperature} dependence of the
dynamics (FSC) and thermodynamics (inflection in $\langle e(T)
\rangle$, related to the Kauzmann paradox) for BKS99. Going beyond
previous work the main goal of this work is to analyse the
relevant observables in terms of {\it energy}. From this we will
uniquely identify a common origin of the FSC and the inflection in
$\langle e(T) \rangle$: just around $T_{FSC}$ the system
approaches the low-energy end of the PEL.

One of the most fundamental energy-dependent quantities is the
density of MBs $G(e)$. In a first step we record the distribution
of MBs $p(e,T)$ at different temperatures. In case that
anharmonicities are small or do not strongly depend on temperature
one has $p(e,T) \propto G(e)\exp(-\beta e)$ with $\beta =
1/(k_BT)$ or, equivalently, $G(e) \propto p(e,T) \exp(\beta e)$
\cite{Sciortino:1999,Buechner:1999}. Thus via reweighting it is
possible to extract $G(e)$ from $p(e,T)$ apart from a
proportionality constant. Actually, for $T > 5000$ K this
reweighting breaks down because of anharmonicities.  We have
estimated the proportionality constant from the condition that for
$T = 5000$ K the configurational entropy is identical to the value
of 6.8 J/(mol K), reported in \cite{Voivod2:cond}. The resulting
energy-distribution $G(e)$ is shown in Fig.2. It is possible to
fit $G(e)$ with a gaussian $G(e) = c
\exp(-(e-e_0)^2/(2\sigma^2))/\sqrt{2\pi \sigma^2}$ with $c =
10^{49}, \sigma^2 = 12$ eV$^2$, and $e_0 = -1867.2$ eV. In analogy
to the binary Lennard-Jones system the distribution is gaussian
within statistical errors.

\begin{figure}
\includegraphics[width=8.6cm]{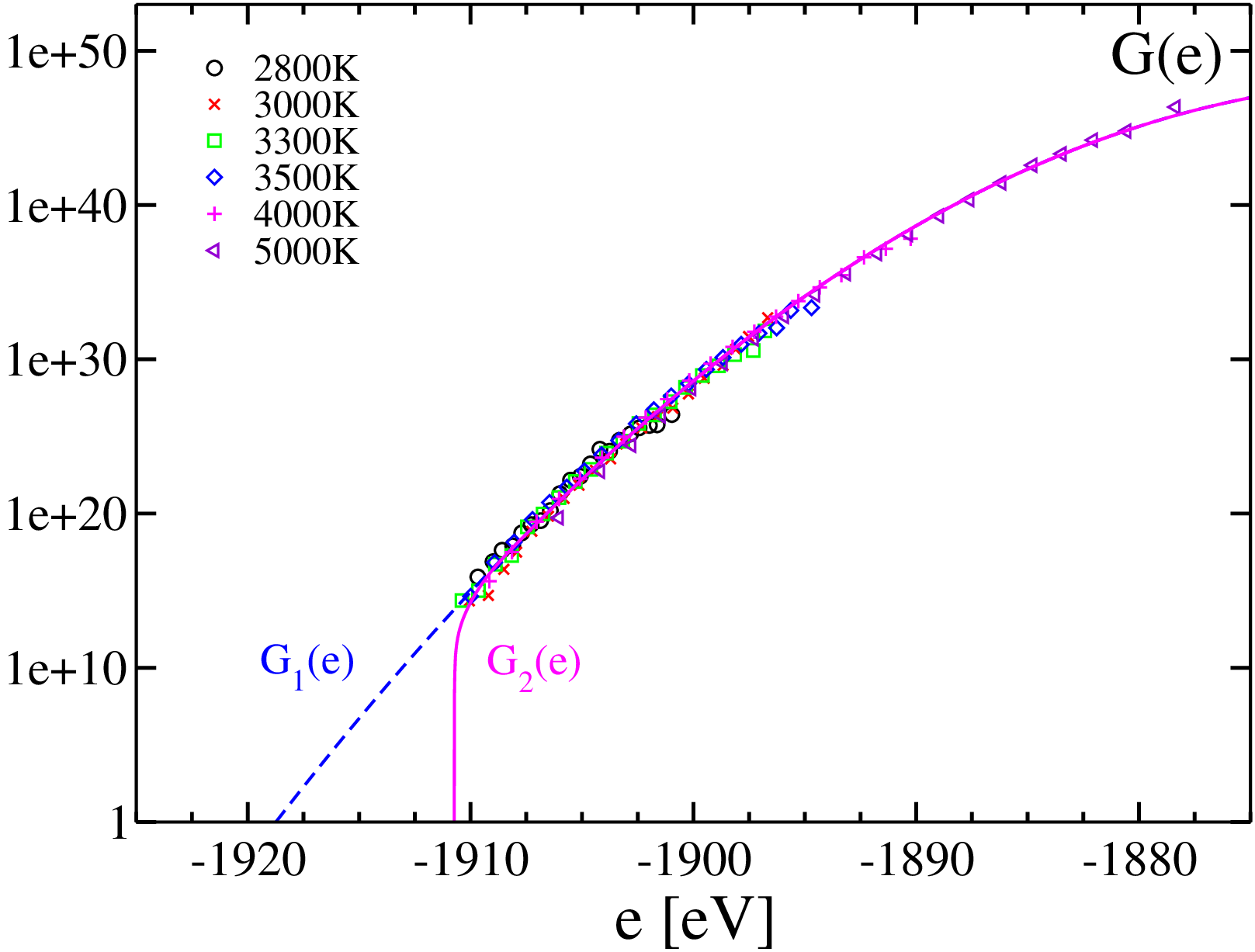}
\caption{\label{Ge} The energy distribution $G(e)$, obtained from
reweighting of $p(e,T)$ for temperatures between 2800 K and 5000
K. The resulting gaussian-type fits $G_1(e)$ and $G_2(e)$ are
included (see text).}
\end{figure}

Naturally, during our simulations we find a lowest MB energy,
which we denote as $e_c$ ($e_c = -1910.7$ eV). In principle there
are two possible scenarios for this cutoff: (1) The density of
states $G(e)$ is also Gaussian for $e < e_c$. These states,
however, are not found due to finite simulation times. The
resulting distribution is denoted $G_1(e)$. (2) The number of
states below $e_c$ is much smaller than predicted by the Gaussian
extrapolation such that $e_c$ can be basically regarded as a
low-energy cutoff of the energy landscape (denoted
$G_{cutoff}(e))$. To falsify scenario (1) we assume the validity
of $G_1(e)$. In a first step one can compare the average energy
with the gaussian prediction $\langle e(T) \rangle = e_0 -
\sigma^2/T$ \cite{Sciortino:1999,Buechner:1999}. It is included in
Fig.1b. One can clearly see the deviations for $T < 4000$ K,
indicating that the description by a gaussian breaks down at low
energies. Actually, the deviations at high temperatures are due to
anharmonicities.
 An even closer analysis can be
performed by comparing the whole distribution $p(e,T)$, obtained
from our simulations, with $p_1(e,T) \propto G_1(e) \exp(-\beta
e)$. This comparison is shown in Fig.3. The dramatic difference
between $p_1(e,T$=2800 K$)$ and $p(e,T$=2800 K$)$ again shows that
the gaussian extrapolation is not valid. Actually, this difference
can be also expressed by stating that more than 100 MBs with
energies below $e_c$ should have been found during our simulation
time of 120 ns at $T=2800$ K in case of scenario (1). This
quantification, however, is rather involved. Since the discrepancy
is already obvious from Fig.3 we defer it to a later publication.
In any event, we have to conclude that the appearance of a cutoff
is not an artifact but reflects a major depletion of states below
$e_c$ as compared to the gaussian distribution. Actually, using the classification scheme 
recently introduced by 
Ruocco et al. this result clearly shows that BKS silica is a "B strong liquid" \cite{Ruocco:2004}.

Before validating scenario (2) we slightly smear out the cutoff by
introducing the distribution $G_2(e) \propto G_{cutoff}(e) \cdot
\tanh((e-e_c)^2/2)$, defined for $e
> e_c$ (see Fig.\ref{Ge}). As shown in Fig.3
the estimation $p_2(e,T) \propto G_2(e) \exp(-\beta e)$ reproduces
$p(e,T)$ for all temperatures.

This analysis may help to clarify why it is so important to use
rather small systems. Based on the assumption that a system of
$n\cdot 99$ ($n=2,3$) particles is composed of $n$ independent but
identical subsets of 99 particles it possible to estimate
$p(e/n,T=2800 $ K); see inset of Fig.3. Whereas for 99 particles
the cutoff can be directly seen from the strongly asymmetric shape
of $p(e,T)$, this effect is smeared out for 198 or 297 particles
and in the limit of large $n$, $p(e/n,T)$ would finally approach a
delta-function. Thus the presence of the cutoff in any subsystem
is no longer visible for larger systems. This line of thought
explicitly rationalizes the use of small systems.

\begin{figure}
\includegraphics[width=8.6cm]{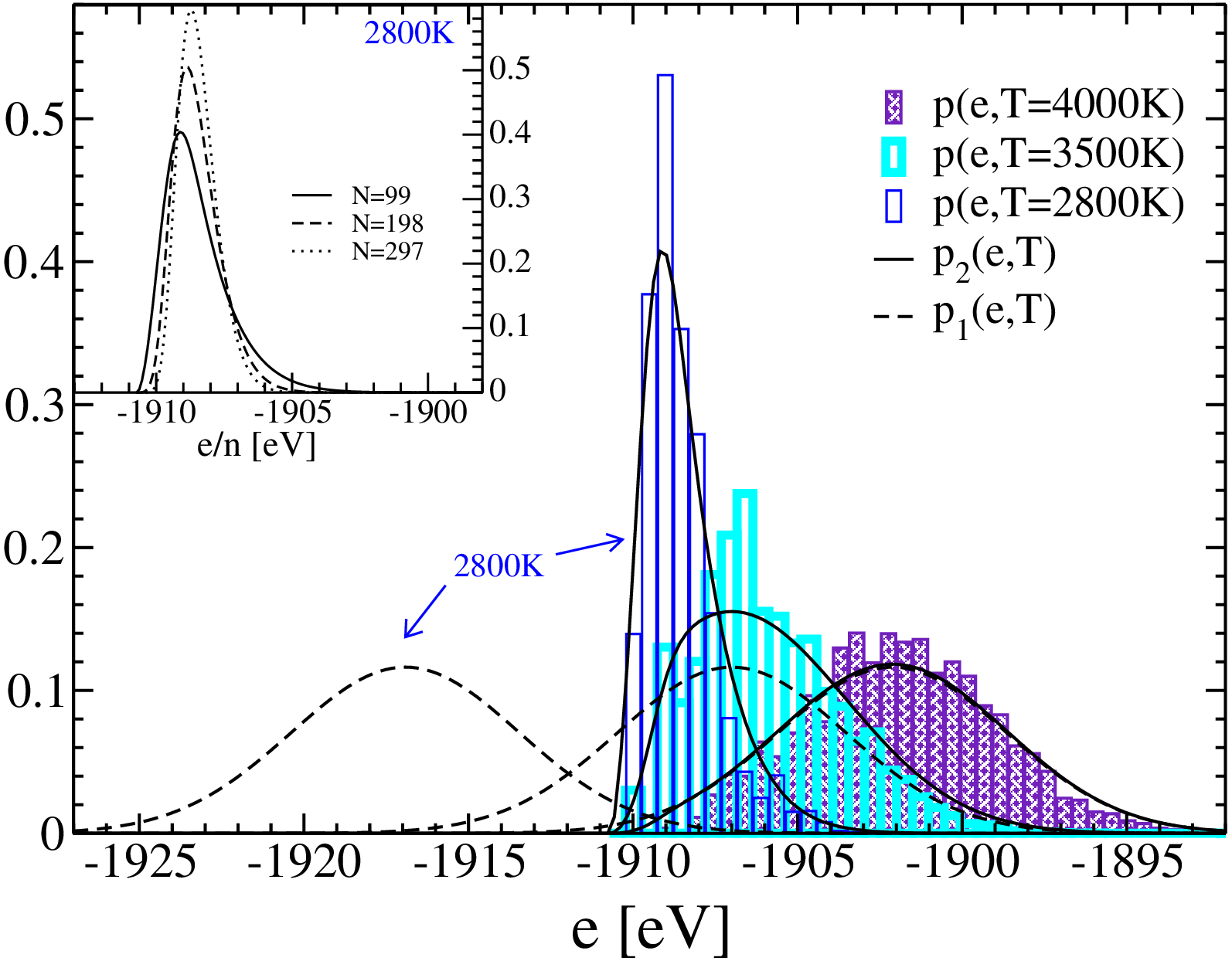}
\caption{\label{d2eff} The simulated distribution $p(e,T)$ for
T=2800 K, T=3500 K, and T=4000 K. Furthermore we include $p_1(e)$
and $p_2(e,T)$ (for the same three temperatures). Note that
$p_1(e,T=4000$ K) $ \approx p_2(e,T=4000$ K). In the inset the
estimation of $p_2(e/n,T=2800K)$ for different systems sizes
$n\cdot 99$ (n=1,2,3) is displayed.}
\end{figure}

Whereas for high temperatures the thermodynamics and the dynamics
is not influenced by the cutoff at $e_c$, a significant impact is
expected at low temperatures. As seen from Fig.3 for temperatures
below 4000 K the cutoff-population $p(e \approx e_c,T)$ starts to
become relevant, resulting in an increasing difference between
$p_1(e,T)$ and $p(e,T)$.  This is exactly the temperature range of
the FSC. We suggest that this agreement is not coincidental. As
shown for the more fragile binary Lennard-Jones system the
increase of the apparent activation energy with decreasing
temperature can be related to the fact that the system explores
lower regions of the PEL at low temperatures for which the local
effective saddles to escape the MB are correspondingly higher
\cite{Doliwa_hop:2003}. Not surprisingly, the same phenomenon is
seen for BKS99 (data not shown). Thus the growing influence of the
bottom of the PEL with decreasing temperature stops the increase
of the apparent activation energy.

\begin{figure}
\includegraphics[width=8.6cm]{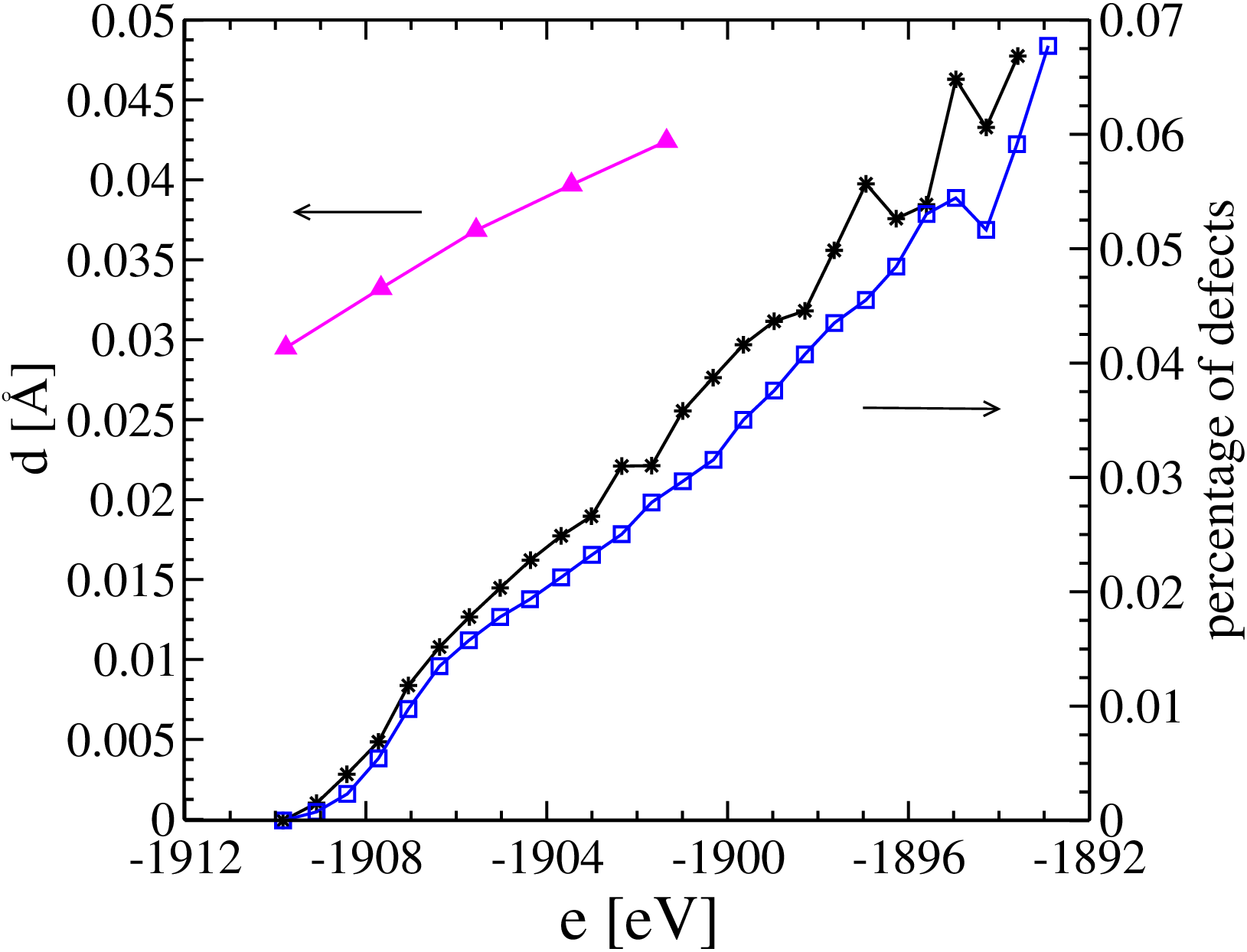}
\caption{\label{d2eff} The fraction of silicon (oxygen) atoms with
3 or 5 oxygen (1 or 3 silicon) atoms in the nearest neighbor shell
for different temperatures (right axis; silicon: stars; oxygen:
squares) and the width of the partial Si-O structure factor for
the nearest-neighbor shell (left axis). }
\end{figure}

What is the microscopic origin of the low-energy cutoff?  One may
expect that the energy of a structure is correlated with the
number of defects, i.e. silicon atoms with three or five oxygens
in the nearest neighbor shell or oxygens with one or three silicon
atoms. This is suggested by the significant decrease of the number
of defects with decreasing temperature, observed in previous
simulations \cite{Horbach:1999,Voivod:2004}. In a first step, we
compare the occurrence of defects with the results for BKS8000
\cite{Horbach:1999}. It turns out that the number of different
defects of the equilibrium configurations at $T = 4000$ K on
average agree within a factor of 1.15 with the numbers, reported
in \cite{Horbach:1999}. Thus the properties of defects do not show
major finite size effects. In a second step we determine the
fraction of these defects (nearest neighbor-shell: $r_{SiO} <
2.28$ \AA) for the minimized structures in dependence of energy;
see Fig.4. One can clearly see that the fraction of defects varies
roughly linearly with energy and approaches zero for
configurations around $e \approx e_c$. These configurations,
however, are still amorphous. This is checked by calculating the
width of the first nearest-neighbor peak of the Si-O partial
structure factor for the minimized structures. It is a measure for
the degree of disorder in the nearest-neighbor shell. This value
only weakly depends on energy; see Fig.4. We may conclude that the
reduction of the number of defects is an efficient way to obtain
configurations with lower energy. After reaching configurations
with no defects at all there emerges a lower energy limit which
cannot be crossed without crystallisation. This suggests a
structural reason for this cutoff. In particular this excludes the
scenario that the cutoff is a trivial consequence of the fact that
for a finite number of MBs (or inherent structures) at some
temperature the lowest energy configuration becomes
thermodynamically relevant. As seen in Fig.2 we indeed estimate an
exponentially large number of different states around $e \approx
e_c$.

This scenario is very different as compared to the fragile binary
Lennard-Jones system where $p_1(e,T) \approx p(e,T)$ for all
simulated temperatures \cite{Doliwa_tau:2003}, excluding a PEL
cutoff in the accessible energy range. Without network constraints
the system can optimize the disordered structure more efficiently
to find configurations with even lower energies (see also
\cite{Voivod:2004}).

Is it possible that the FSC for macroscopic BKS is of different
nature than for BKS99? One might think that for larger systems it
is possible to form more low-energy states due to the smaller
number of constraints related to the periodic boundary conditions.
Then $G(e)$ might extend the gaussian-like behaviour also to lower
energies and the temperature dependence of $\langle e(T) \rangle $
should approach the gaussian limit. Comparison of BKS99 with
BKS1332 shows that this effect, if at all, is very small (Fig.1).
Therefore macroscopic BKS not only displays the same phenomenology
as BKS99 but very likely also follows similar underlying
mechanisms.

In summary, for the prototype network former silica we have
related the fragile-to-strong crossover as well as the inflection
of $\langle e(T) \rangle$  to the presence of a lower cutoff in
the PEL. Since this cutoff is related to structural constraints
for disordered structures of a network former, it may be of
general nature.

We would like to acknowledge discussions with B. Doliwa and M.
Vogel. We thank F. Sciortino and I. Saika-Voivod for sending us
the data of Ref.\cite{Voivod2:cond}.

\bibliography{PELsio2}

\begin{thebibliography}{35}
\expandafter\ifx\csname natexlab\endcsname\relax\def\natexlab#1{#1}\fi
\expandafter\ifx\csname bibnamefont\endcsname\relax
  \def\bibnamefont#1{#1}\fi
\expandafter\ifx\csname bibfnamefont\endcsname\relax
  \def\bibfnamefont#1{#1}\fi
\expandafter\ifx\csname citenamefont\endcsname\relax
  \def\citenamefont#1{#1}\fi
\expandafter\ifx\csname url\endcsname\relax
  \def\url#1{\texttt{#1}}\fi
\expandafter\ifx\csname urlprefix\endcsname\relax\def\urlprefix{URL }\fi
\providecommand{\bibinfo}[2]{#2}
\providecommand{\eprint}[2][]{\url{#2}}

\bibitem[{\citenamefont{{C. A. Angell}}(1991)}]{Angell:1991}
\bibinfo{author}{\bibnamefont{{C. A. Angell}}}, \bibinfo{journal}{J. Non-Cryst.
  Solids.} \textbf{\bibinfo{volume}{131-133}}, \bibinfo{pages}{13 }
  (\bibinfo{year}{1991}).

\bibitem[{\citenamefont{{K. Vollmayr, }{W. Kob, }{and K.
  Binder}}(1996)}]{Vollmayr:1996}
\bibinfo{author}{\bibnamefont{{K. Vollmayr, }{W. Kob, }{and K. Binder}}},
  \bibinfo{journal}{Phys. Rev. B} \textbf{\bibinfo{volume}{54}},
  \bibinfo{pages}{15808} (\bibinfo{year}{1996}).

\bibitem[{\citenamefont{{I. Saika-Voivod, }{F. Sciortino, }{and P. H.
  Poole}}(2001{\natexlab{a}})}]{Voivod:2001}
\bibinfo{author}{\bibnamefont{{I. Saika-Voivod, }{F. Sciortino, }{and P. H.
  Poole}}}, \bibinfo{journal}{Phys. Rev. E} \textbf{\bibinfo{volume}{63}},
  \bibinfo{pages}{011202} (\bibinfo{year}{2001}{\natexlab{a}}).

\bibitem[{\citenamefont{{M. S. Shell, }{P. G. Debenedetti, }{and A. Z.
  Panagiotopoulos}}(2002)}]{Shell:2002}
\bibinfo{author}{\bibnamefont{{M. S. Shell, }{P. G. Debenedetti, }{and A. Z.
  Panagiotopoulos}}}, \bibinfo{journal}{Phys. Rev. E}
  \textbf{\bibinfo{volume}{66}}, \bibinfo{pages}{011202}
  (\bibinfo{year}{2002}).

\bibitem[{\citenamefont{{J. C. Mikkelsen}}(1984)}]{Mikkelsen:1984}
\bibinfo{author}{\bibnamefont{{J. C. Mikkelsen}}}, \bibinfo{journal}{Appl.
  Phys. Lett.} \textbf{\bibinfo{volume}{45}}, \bibinfo{pages}{1187 }
  (\bibinfo{year}{1984}).

\bibitem[{\citenamefont{{K.-U. Hess, }{D. B. Dingwell, }{and E.
  R\"ossler}}(1996)}]{Hess:1996}
\bibinfo{author}{\bibnamefont{{K.-U. Hess, }{D. B. Dingwell, }{and E.
  R\"ossler}}}, \bibinfo{journal}{Chem. Geol.} \textbf{\bibinfo{volume}{128}},
  \bibinfo{pages}{155} (\bibinfo{year}{1996}).

\bibitem[{\citenamefont{{J. Horbach}{ and W. Kob}}(1999)}]{Horbach:1999}
\bibinfo{author}{\bibnamefont{{J. Horbach}{ and W. Kob}}},
  \bibinfo{journal}{Phys. Rev. B.} \textbf{\bibinfo{volume}{60}},
  \bibinfo{pages}{3169 } (\bibinfo{year}{1999}).

\bibitem[{\citenamefont{{C. A. Angell }}(1983)}]{Angell:1983}
\bibinfo{author}{\bibnamefont{{C. A. Angell }}}, \bibinfo{journal}{Annu. Rev.
  Phys. Chem.} \textbf{\bibinfo{volume}{34}}, \bibinfo{pages}{7079}
  (\bibinfo{year}{1983}).

\bibitem[{\citenamefont{{M. Hemmati, }{C. T. Moynihan, }{and C. A.
  Angell}}(2001)}]{Hemmati:2001}
\bibinfo{author}{\bibnamefont{{M. Hemmati, }{C. T. Moynihan, }{and C. A.
  Angell}}}, \bibinfo{journal}{J. Chem. Phys.} \textbf{\bibinfo{volume}{115}},
  \bibinfo{pages}{6663} (\bibinfo{year}{2001}).

\bibitem[{\citenamefont{{F. W. Starr, }{F. Sciortino, }{and H. E.
  Stanley}}(1999)}]{Starr_Sciortino:1999}
\bibinfo{author}{\bibnamefont{{F. W. Starr, }{F. Sciortino, }{and H. E.
  Stanley}}}, \bibinfo{journal}{Phys. Rev. E.} \textbf{\bibinfo{volume}{60}},
  \bibinfo{pages}{6757 } (\bibinfo{year}{1999}).

\bibitem[{\citenamefont{{I. Saika-Voivod, }{P. H. Poole, }{and F.
  Sciortino}}(2003)}]{Voivod2:cond}
\bibinfo{author}{\bibnamefont{{I. Saika-Voivod, }{P. H. Poole, }{and F.
  Sciortino}}}, \bibinfo{journal}{arXiv:cond-mat/0309481}
  (\bibinfo{year}{2003}).

\bibitem[{\citenamefont{{F. H. Stillinger }{and T. A.
  Weber}}(1982)}]{Stillinger1:1982}
\bibinfo{author}{\bibnamefont{{F. H. Stillinger }{and T. A. Weber}}},
  \bibinfo{journal}{Phys. Rev. A} \textbf{\bibinfo{volume}{25}},
  \bibinfo{pages}{978} (\bibinfo{year}{1982}).

\bibitem[{\citenamefont{{S. Sastry, }{P. G. Debenedetti, }{and F. H.
  Stillinger}}(1998)}]{Sastry:1998}
\bibinfo{author}{\bibnamefont{{S. Sastry, }{P. G. Debenedetti, }{and F. H.
  Stillinger}}}, \bibinfo{journal}{Nature} \textbf{\bibinfo{volume}{393}},
  \bibinfo{pages}{554} (\bibinfo{year}{1998}).

\bibitem[{\citenamefont{{F. Sciortino, }{W. Kob, }{and P.
  Tartaglia}}(1999)}]{Sciortino:1999}
\bibinfo{author}{\bibnamefont{{F. Sciortino, }{W. Kob, }{and P. Tartaglia}}},
  \bibinfo{journal}{Phys. Rev. Lett.} \textbf{\bibinfo{volume}{83}},
  \bibinfo{pages}{3214} (\bibinfo{year}{1999}).

\bibitem[{\citenamefont{{S. B\"{u}chner }{and A. Heuer}}(1999)}]{Buechner:1999}
\bibinfo{author}{\bibnamefont{{S. B\"{u}chner }{and A. Heuer}}},
  \bibinfo{journal}{Phys. Rev. E} \textbf{\bibinfo{volume}{60}},
  \bibinfo{pages}{6507} (\bibinfo{year}{1999}).

\bibitem[{\citenamefont{{A. Scala, }{F. W. Starr, }{E. La Nave, }{F. Sciortino,
  }{and H. E. Stanley}}(2000)}]{Scala:2000}
\bibinfo{author}{\bibnamefont{{A. Scala, }{F. W. Starr, }{E. La Nave, }{F.
  Sciortino, }{and H. E. Stanley}}}, \bibinfo{journal}{Nature}
  \textbf{\bibinfo{volume}{406}}, \bibinfo{pages}{166} (\bibinfo{year}{2000}).

\bibitem[{\citenamefont{{P. G. Debenedetti}{ and F. H.
  Stillinger}}(2001)}]{Debenedetti:2001}
\bibinfo{author}{\bibnamefont{{P. G. Debenedetti}{ and F. H. Stillinger}}},
  \bibinfo{journal}{Nature} \textbf{\bibinfo{volume}{410}}, \bibinfo{pages}{259
  } (\bibinfo{year}{2001}).

\bibitem[{\citenamefont{{D. J. Wales }{and J. P. K. Doye}}(2001)}]{Wales:2001}
\bibinfo{author}{\bibnamefont{{D. J. Wales }{and J. P. K. Doye}}},
  \bibinfo{journal}{Phys. Rev. B.} \textbf{\bibinfo{volume}{63}},
  \bibinfo{pages}{214204 } (\bibinfo{year}{2001}).

\bibitem[{\citenamefont{{S. Sastry}}(2001)}]{Sastry:2001}
\bibinfo{author}{\bibnamefont{{S. Sastry}}}, \bibinfo{journal}{Nature}
  \textbf{\bibinfo{volume}{409}}, \bibinfo{pages}{164} (\bibinfo{year}{2001}).

\bibitem[{\citenamefont{{E. La Nave, }{H. E. Stanley,}{ and F.
  Sciortino}}(2002)}]{Nave_Sciortino:2002}
\bibinfo{author}{\bibnamefont{{E. La Nave, }{H. E. Stanley,}{ and F.
  Sciortino}}}, \bibinfo{journal}{Phys. Rev. Lett.}
  \textbf{\bibinfo{volume}{88}}, \bibinfo{pages}{035501 }
  (\bibinfo{year}{2002}).

\bibitem[{\citenamefont{{M. S. Shell }{and P. G.
  Debenedetti}}(2003)}]{Shell:2003}
\bibinfo{author}{\bibnamefont{{M. S. Shell }{and P. G. Debenedetti}}},
  \bibinfo{journal}{J. Chem. Phys.} \textbf{\bibinfo{volume}{118}}
  (\bibinfo{year}{2003}).

\bibitem[{\citenamefont{{B. Doliwa }{and A.
  Heuer}}(2003{\natexlab{a}})}]{Doliwa_hop:2003}
\bibinfo{author}{\bibnamefont{{B. Doliwa }{and A. Heuer}}},
  \bibinfo{journal}{Phys. Rev. E.} \textbf{\bibinfo{volume}{67}},
  \bibinfo{pages}{030501} (\bibinfo{year}{2003}{\natexlab{a}}).

\bibitem[{\citenamefont{{I. Saika-Voivod, }{P. H. Poole, }{and F.
  Sciortino}}(2001)}]{Voivod1:2001}
\bibinfo{author}{\bibnamefont{{I. Saika-Voivod, }{P. H. Poole, }{and F.
  Sciortino}}}, \bibinfo{journal}{Nature} \textbf{\bibinfo{volume}{412}},
  \bibinfo{pages}{514} (\bibinfo{year}{2001}).

\bibitem[{\citenamefont{{I. Saika-Voivod, }{F. Sciortino, }{and P. H.
  Poole}}(2001{\natexlab{b}})}]{Voivod:2000}
\bibinfo{author}{\bibnamefont{{I. Saika-Voivod, }{F. Sciortino, }{and P. H.
  Poole}}}, \bibinfo{journal}{Phys. Rev. E} \textbf{\bibinfo{volume}{63}},
  \bibinfo{pages}{011202} (\bibinfo{year}{2001}{\natexlab{b}}).

\bibitem[{\citenamefont{{B. Doliwa }{and A.
  Heuer}}(2003{\natexlab{b}})}]{Doliwa_tau:2003}
\bibinfo{author}{\bibnamefont{{B. Doliwa }{and A. Heuer}}},
  \bibinfo{journal}{Phys. Rev. E.} \textbf{\bibinfo{volume}{67}},
  \bibinfo{pages}{031506} (\bibinfo{year}{2003}{\natexlab{b}}).

\bibitem[{\citenamefont{{B. W. H. van Beest, }{G. J. Kramer, }{and R. A. van
  Santen}}(1990)}]{BKS:1990}
\bibinfo{author}{\bibnamefont{{B. W. H. van Beest, }{G. J. Kramer, }{and R. A.
  van Santen}}}, \bibinfo{journal}{Phys. Rev. Lett.}
  \textbf{\bibinfo{volume}{64}}, \bibinfo{pages}{1955} (\bibinfo{year}{1990}).

\bibitem[{\citenamefont{{F. H. Stillinger}}(1995)}]{Stillinger:1995}
\bibinfo{author}{\bibnamefont{{F. H. Stillinger}}}, \bibinfo{journal}{Science}
  \textbf{\bibinfo{volume}{267}}, \bibinfo{pages}{1935} (\bibinfo{year}{1995}).

\bibitem[{\citenamefont{{S. Buechner }{and A. Heuer}}(2000)}]{Buechner:2000}
\bibinfo{author}{\bibnamefont{{S. Buechner }{and A. Heuer}}},
  \bibinfo{journal}{Phys. Rev. Lett.} \textbf{\bibinfo{volume}{84}},
  \bibinfo{pages}{2168} (\bibinfo{year}{2000}).

\bibitem[{\citenamefont{{T. F. Middleton }{and D. J.
  Wales}}(2001)}]{Middleton:2001}
\bibinfo{author}{\bibnamefont{{T. F. Middleton }{and D. J. Wales}}},
  \bibinfo{journal}{Phys. Rev. B} \textbf{\bibinfo{volume}{64}},
  \bibinfo{pages}{024205} (\bibinfo{year}{2001}).

\bibitem[{\citenamefont{{D. J. Wales }{and J. P. K. Doye}}(2003)}]{Wales:2003}
\bibinfo{author}{\bibnamefont{{D. J. Wales }{and J. P. K. Doye}}},
  \bibinfo{journal}{J. Chem. Phys.} \textbf{\bibinfo{volume}{119}},
  \bibinfo{pages}{12409} (\bibinfo{year}{2003}).

\bibitem[{\citenamefont{{B. Doliwa }{and A.
  Heuer}}(2003{\natexlab{c}})}]{Doliwa_finite:2003}
\bibinfo{author}{\bibnamefont{{B. Doliwa }{and A. Heuer}}},
  \bibinfo{journal}{J. Phys. C: Cond. Mat.} \textbf{\bibinfo{volume}{15}},
  \bibinfo{pages}{S849} (\bibinfo{year}{2003}{\natexlab{c}}).

\bibitem[{\citenamefont{{J. Horbach,}{ W. Kob,}{ K. Binder,}{ and C. A.
  Angell}}(1996)}]{Horbach_finite:1996}
\bibinfo{author}{\bibnamefont{{J. Horbach,}{ W. Kob,}{ K. Binder,}{ and C. A.
  Angell}}}, \bibinfo{journal}{Phys. Rev. E.} \textbf{\bibinfo{volume}{54}},
  \bibinfo{pages}{R5897 } (\bibinfo{year}{1996}).

\bibitem[{\citenamefont{{M. Vogel }{and S. C. Glotzer}}(2004)}]{Vogel:2004}
\bibinfo{author}{\bibnamefont{{M. Vogel }{and S. C. Glotzer}}},
  \bibinfo{journal}{arXiv:cond-mat/0402427}  (\bibinfo{year}{2004}).

\bibitem[{\citenamefont{{G. Ruocco, }{F. Sciortino, }{F. Zamponi, }{C. De
  Michele}{and T. Scopigno}}(2004)}]{Ruocco:2004}
\bibinfo{author}{\bibnamefont{{G. Ruocco, }{F. Sciortino, }{F. Zamponi, }{C. De
  Michele}{and T. Scopigno}}}, \bibinfo{journal}{arXiv:cond-mat/0401449}
  (\bibinfo{year}{2004}).

\bibitem[{\citenamefont{{I. Saika-Voivod, }{P. H. Poole, }{and F.
  Sciortino}}(2004)}]{Voivod:2004}
\bibinfo{author}{\bibnamefont{{I. Saika-Voivod, }{P. H. Poole, }{and F.
  Sciortino}}}, \bibinfo{journal}{Phil. Mag.} \textbf{\bibinfo{volume}{48}},
  \bibinfo{pages}{1437} (\bibinfo{year}{2004}).

\end{thebibliography}

\end{document}